\documentclass[aps,pre,twocolumn,superscriptaddress,groupedaddress]{revtex4}  
\usepackage{graphicx}  
\usepackage{url}                  
\usepackage{dcolumn}       
\usepackage{bm}                  
\usepackage{amssymb}       
\usepackage{mathtools}     
\DeclarePairedDelimiter\ket{\lvert}{\rangle}
\usepackage{amsmath}    
\usepackage{pgf}                   
\usepackage{tikz,pgfplots}                   
\usetikzlibrary{shapes,arrows,calc}
\usetikzlibrary{arrows, decorations.markings}
\usepackage{lipsum}

\usepackage{soul}

\newcommand{\Tmean}[1]{\langle  #1 \rangle_{t}}                 

\definecolor{orange}{rgb}{0.8, 0.4, 0.2}

\begin{document}  

\title{Reduced-order modeling of two-dimensional turbulent Rayleigh-B\'{e}nard flow by hybrid quantum-classical reservoir computing}
\author{Philipp Pfeffer}
\affiliation{Institut f\"ur Thermo- und Fluiddynamik, Technische Universit\"at Ilmenau, Postfach 100565, D-98684 Ilmenau, Germany}
\author{Florian Heyder}
\affiliation{Institut f\"ur Thermo- und Fluiddynamik, Technische Universit\"at Ilmenau, Postfach 100565, D-98684 Ilmenau, Germany}
\author{J\"org Schumacher}
\affiliation{Institut f\"ur Thermo- und Fluiddynamik, Technische Universit\"at Ilmenau, Postfach 100565, D-98684 Ilmenau, Germany}
\affiliation{Tandon School of Engineering, New York University, New York City, NY 11201, USA}
\date{\today}

\begin{abstract}
Two hybrid quantum-classical reservoir computing models are presented to reproduce low-order statistical properties of a two-dimensional turbulent Rayleigh-B\'{e}nard convection flow at a Rayleigh number $Ra=10^5$ and a Prandtl number $Pr=10$. These properties comprise the mean vertical profiles of the root mean square velocity and temperature and the turbulent convective heat flux. The latter is composed of vertical velocity and temperature and measures the global turbulent heat transfer across the convection layer; it manifests locally in coherent hot and cold thermal plumes that rise from the bottom and fall from the top boundaries. Both quantum algorithms differ by the arrangement of the circuit layers of the quantum reservoir, in particular the entanglement layers. The second of the two quantum circuit architectures, denoted as H2, enables a complete execution of the reservoir update inside the quantum circuit without the usage of external memory. Their performance is compared with that of a classical reservoir computing model. Therefore, all three models have to learn the nonlinear and chaotic dynamics of the turbulent flow at hand in a lower-dimensional latent data space which is spanned by the time-dependent expansion coefficients of the 16 most energetic Proper Orthogonal Decomposition (POD) modes. These training data are generated by a POD snapshot analysis from direct numerical simulations of the original turbulent flow. All reservoir computing models are operated in the reconstruction or open-loop mode, i.e., they receive 3 POD modes as an input at each step and reconstruct the missing 13 ones. We analyse different measures of the reconstruction error in dependence on the hyperparameters which are specific for the quantum cases or shared with the classical counterpart, such as the reservoir size and the leaking rate. We show that both quantum algorithms are able to reconstruct the essential statistical properties of the turbulent convection flow successfully with similar performance compared to the classical reservoir network. Most importantly, the quantum reservoirs are by a factor of 4 to 8 smaller in comparison to the classical case.      
\end{abstract}
\maketitle

\section{Introduction}
Quantum computing algorithms have changed our ways to process, classify, generate, and analyse data \cite{Preskill2018,Deutsch2020}. New ways to solve classical fluid mechanical problems have been suggested in the form of quantum amplitude estimation algorithms \cite{Gaitan2020}, variational quantum algorithms \cite{Lubasch2020,Kyriienko2021}, quantum Lattice Boltzmann methods \cite{Todorova2020,Budinski2021,Schalkers2022,Li2023,Itani2023}, or quantum linear system algorithms \cite{Bharadwaj2020,Liu2021,Bharadwaj2023} for one-dimensional problems. Fluid equations for inviscid or viscous fluids have been also transformed into Schrödinger equations for specific potentials \cite{Succi2023,Meng2023}. Following ref. \cite{Floether2023}, the applications of quantum computing can be grouped into three major fields: (1) simulation of chemical or physical processes, (2) search and optimization, and (3) processing data with complex structure. The last field comprises quantum machine learning methods \cite{Biamonte2017}, such as quantum generative methods \cite{Lloyd2018,Rudolph2022}, quantum kernel methods \cite{Schuld2019}, and quantum recurrent networks in particular in the form of quantum reservoir computing \cite{Markovic2020,Mujal2021,Mujal2023}. 

In classical reservoir computing, the reservoir is the central building block of the neural network architecture. The reservoir is a sparse random network of neurons that substitutes the batch of successively connected hidden layers of deep convolutional neural networks of other machine learning algorithms \cite{Jaeger2001,Jaeger2004}. It introduces a short-term memory to process sequential data. This is the subject of the present investigation. Here, we will substitute the high-dimensional classical reservoir network by a small parametric quantum circuit in which $n$ qubits span a $2^n$-dimensional complex quantum state space for a highly entangled reservoir state to save memory and computational costs. Quantum reservoir computing can be implemented in two different ways which we describe in brief in the following.

The dynamics of an interacting many-particle quantum system --the quantum reservoir-- is investigated in a so-called {\em analogue} framework. It is characterized by a Hamiltonian $H$ subject to an unitary time evolution promoted by $U(t)$. The time evolution of the density matrix $\rho(t)$, which describes the quantum reservoir state, follows then to
\begin{equation}
\rho(t)=U(t)\rho(0) U^{\dagger}(t) \;\;\mbox{with}\;\; U(t)=\exp\left(-\frac{i}{\hbar}Ht\right)\,.
\label{evolution}
\end{equation}
The operator $H$ is the (many-particle) Hamiltonian and $U^{\dagger}(t)$ the adjoint of $U(t)$. These systems have been implemented in the form of spin ensembles \cite{Fujii2017,Nakajima2019,Kutvonen2020,Sakurai2022}, circuit quantum electrodynamics \cite{Angelatos2021}, arrays of Rydberg atoms \cite{Araiza2022}, single oscillators, and networks of oscillators \cite{Nokkala2021,Govia2021}. They establish a {\em closed quantum system} in the ideal case that follows an ideal unitary time evolution after the input state is prepared. The pure state density matrix $\rho(t)$ is given by the outer product of the (many-particle) state vector $|{\bm \Psi}(t)\rangle$ with itself $\rho(t)=|{\bm \Psi}(t)\rangle\langle{\bm \Psi}(t)|$.
\begin{figure*}
    \includegraphics[scale=0.5]{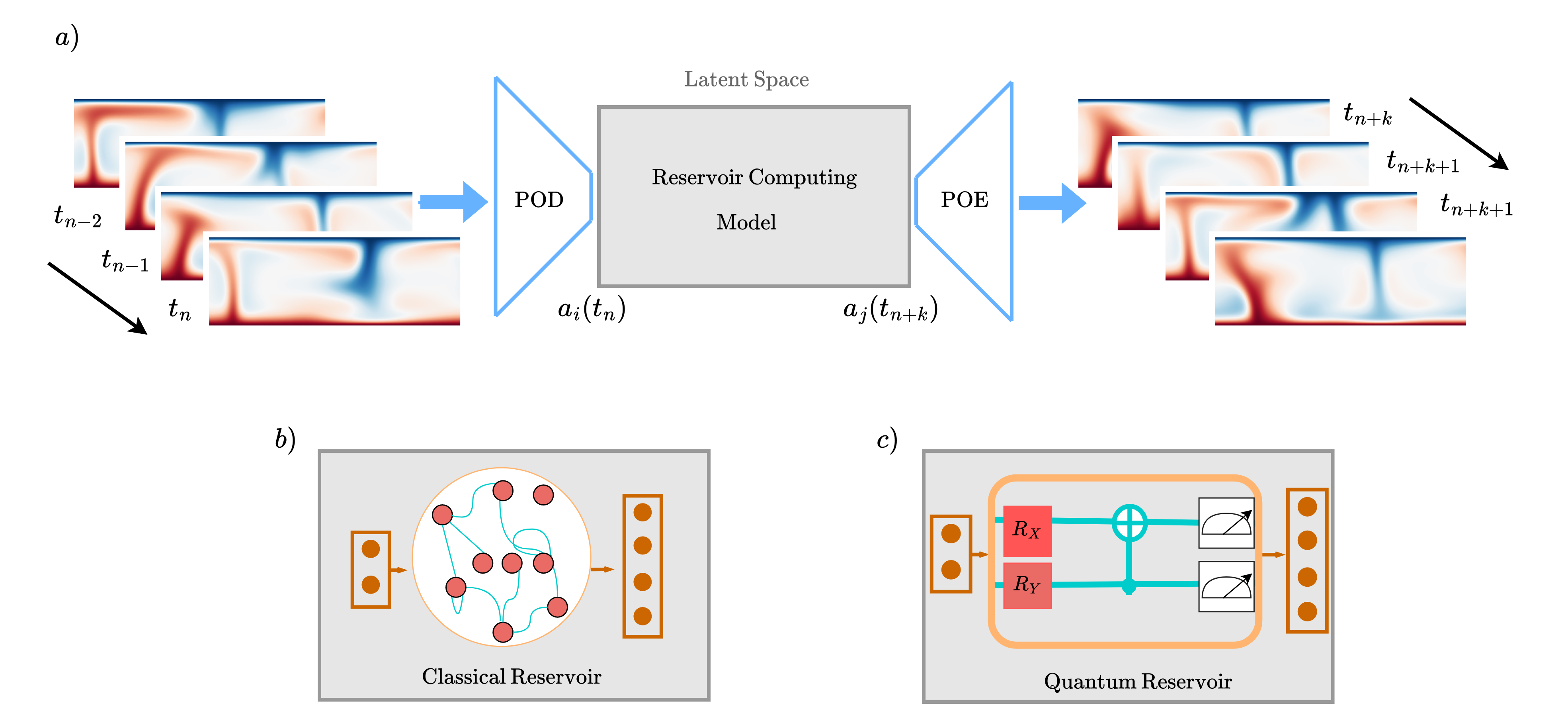}
    \caption{Pipeline of data processing in the classical and hybrid quantum-classical reservoir computing models. (a) Sketch of the whole algorithm. Direct numerical simulation snapshots of the turbulent convection flow to the left are encoded in a low-dimensional latent data space by a Proper Orthogonal Decomposition (POD). A reservoir computing model operates in the latent space receiving the input sequences $a_i(t_n)$ with $i = 1, ..., N_{\rm in}< N_{\rm out}$. Its output is a sequence $a_j(t_{n+k})$ with $j=1, ..., N_{\rm dof}$; typically one takes $k=1$. The output is finally decoded by a Proper Orthogonal Expansion (POE) to simulation data at later times and/or a subsequent statistical analysis. The reservoir takes two different forms in this work. (b) The classical reservoir, which we use for comparison, consists of a sparse random  network of neurons that possesses a dynamic memory for temporal data processing. (c) The quantum reservoir is a digital gate-based quantum circuit. In this work, two different quantum algorithms are employed for the quantum reservoir. Both, classical and quantum reservoir contain input and output layers to the left and right, respectively. The reservoir architectures are further detailed in Sec. IV.}
    \label{fig:pipeline}
\end{figure*}

Beside the analogue framework , the {\em digital gate-based} framework uses parametric circuits composed of universal quantum gates. They are composed to a quantum reservoir on noisy intermediate-scale quantum devices in this case \cite{Chen2020,Suzuki2022}. The reservoir state is obtained by a {\em repeated measurement} of the equivalently prepared quantum system and gives the probabilities $p_j(t)$ for $j=1,\dots,2^n$ of the $n$-qubit quantum state $|{\bm \Psi}(t)\rangle$ to collapse on the $j$-th eigenvector of the standard observable in a quantum computer, the Pauli-$Z$ matrix \cite{Nielsen2010}. These probabilities correspond to the diagonal elements of the density matrix $\rho$ and are summarized to the vector $\bm p(t) = (p_1,\dots,p_N)^{\rm T}=(\rho_{11},\dots,\rho_{NN})^{\rm T}$
with $N=2^n$. They can be red out by a measurement. Again, we assumed that the initial state $\rho(0)$ is a pure state. In ref. \cite{Pfeffer2022}, the digital gate-based approach was realized in the form of an {\em open quantum system} which implies that parts of the short-term memory of the hybrid systems are kept outside the quantum reservoir. This method allowed to model the dynamics of low-dimensional nonlinear systems, such as the Lorenz 63 \cite{Lorenz1963} and its 8-dimensional Lorenz-type model extension \cite{Gluhovsky2002} on an actual IBM quantum computer. This algorithm will be denoted to as hybrid algorithm 1, in short H1, in the following.  

In the present work, we advance our investigation on quantum reservoir computing with a proof-of-concept application to a realistic complex fluid mechanical problem. We seek to show that it is possible to achieve results for the following low-dimensional simulations which are comparable to classical reservoir computing. To this end, we present a full data processing pipeline for a two-dimensional {\em turbulent} Rayleigh-B\'{e}nard convection flow \cite{Chilla2012} which contains a quantum computing module -- the quantum reservoir. This flow is a paradigm for turbulence that is driven by buoyancy forces in many geophysical and astrophysical processes \cite{Stevens2005,Schumacher2020}. The hybrid quantum-classical machine learning algorithm will serve as a data-driven reduced-order model of the turbulent flow without knowledge of the underlying mathematical equations of motion.

The hybrid nature of the quantum machine learning model includes a reduction of the high-dimensional turbulence data to a low-dimensional latent space. This is done by a classical snapshot-based Proper Orthogonal Decomposition (POD) \cite{Sirovich1987}. Similar to classical machine learning algorithms, this encoding/decoding step is necessary for a fully turbulent flow since the dimension of the classical input data is high; the actual number of degrees of freedom is here $\tilde{N}_{\rm dof}=3\times 384\times 96 =110592$, see also \cite{Pandey2022,Heyder2022}. The quantum machine learning algorithm thus operates in a latent data space of $N_{\rm dof}=16$ in the present case and is able to reproduce relevant large-scale features and low-order statistics of the turbulent flow, such as the vertical profile of the mean convective heat flux across the convection layer. Particularly, the latter point is of particular interest in the present application.

We also extend our previous study with an improved hybrid quantum-classical reservoir computing model (RCM) which integrates more parts of the algorithm into the quantum computing part in comparison to the previous algorithm H1 from ref. \cite{Pfeffer2022}. The present work is a first step away from the traditional von Neumann architecture, in which computation and memory are located in distinct components. The new algorithm will be denoted to as H2 in the following. It will be compared in detail with a classical RCM, in short C, and H1, our previous approach.

The hybrid nature of our algorithm implies additionally that the optimization of the reservoir output layer is performed classically by a direct solution of the minimization task.  The full data processing pipeline of the algorithm comprising a combined POD-RCM model is sketched in Fig. \ref{fig:pipeline}. The figure sketches the classical and quantum reservoir in panels (b) and (c). The quantum reservoir builds on a low-qubit-number parametric quantum circuit which spans a high-dimensional reservoir state space based on a highly entangled $n$-qubit quantum state.

The paper is organized as follows. In Sec. II, we present the turbulent flow and in Sec. III the reduction to the low-dimensional latent space in which the quantum reservoir operates. Sec. IV follows with a compact presentation of the algorithms C, H1, and H2. Sec. V discusses the results in dependence on hyperparameters of all three reservoir computing models. Moreover, we also compare different error measures. One is adapted to the specific fluid mechanical application. A summary and an outlook are given in the last section.   

\begin{figure}
\includegraphics[scale=0.3]{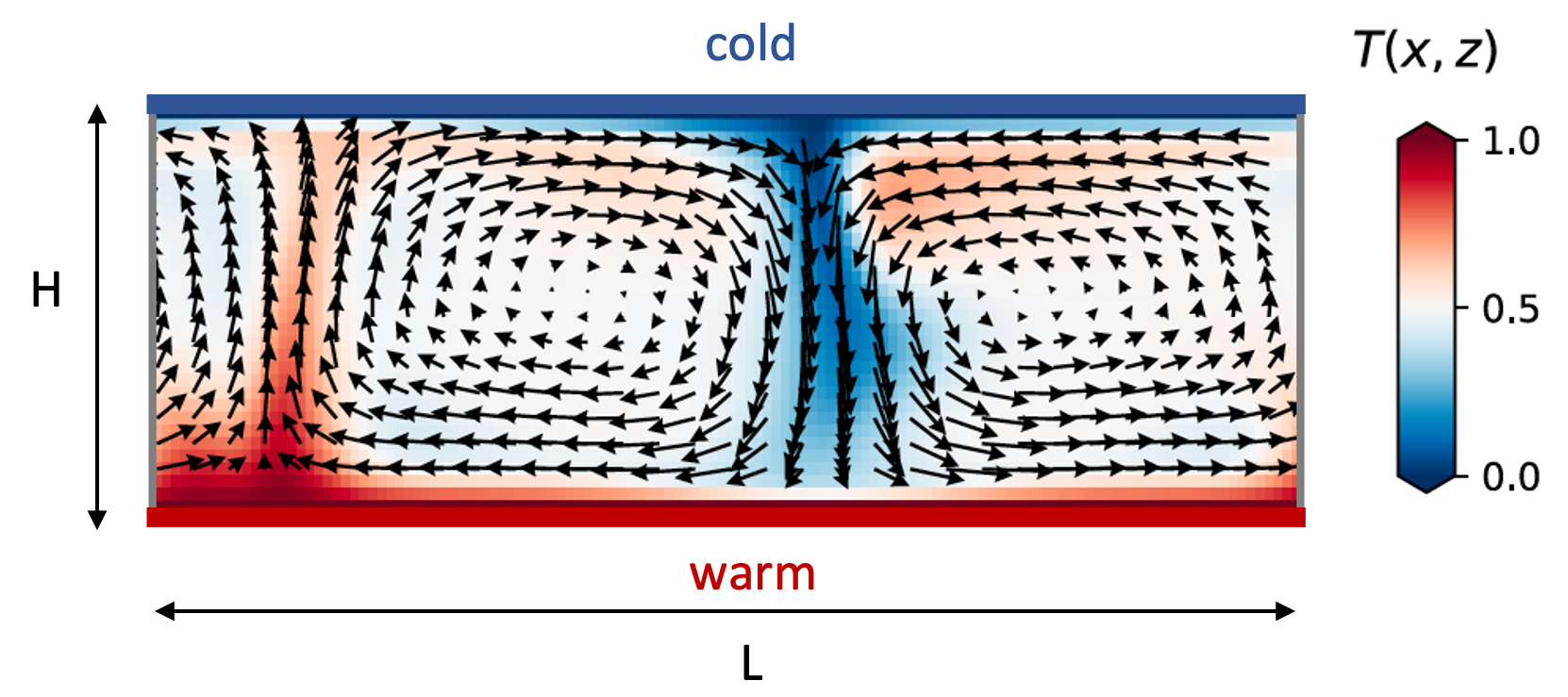}
\caption{Two-dimensional Rayleigh-B\'{e}nard convection flow. The fluid with a kinematic viscosity $\nu$ and temperature diffusivity $\kappa$ is confined between two impermeable boundary planes at the top ($z=H$) and bottom ($z=0$). The bottom plane has a uniform temperature $T_{\rm bot}$; the top one a uniform temperature $T_{\rm top}<T_{\rm bot}$. The fluid layer has a length $L=2\sqrt{2}$ and a height $H=1$, such that the aspect ratio $\Gamma=L/H=2\sqrt{2}$ results. The figure displays a snapshot of the turbulent flow. Background contours stand for the temperature field $T$; the vectors illustrate the 2-component velocity field $(u_x, u_z)$. The turbulent heat transfer across the layer from the bottom to the top is manifest by the thin plume structures which are visible: hot lighter fluid (in red) rises to the top while cold heavier fluid (in blue) falls downwards. Since the velocity field is divergence-free a circulation roll pattern develops.}
\label{fig:RBC}
\end{figure}
\begin{figure*}
\includegraphics[scale=0.5]{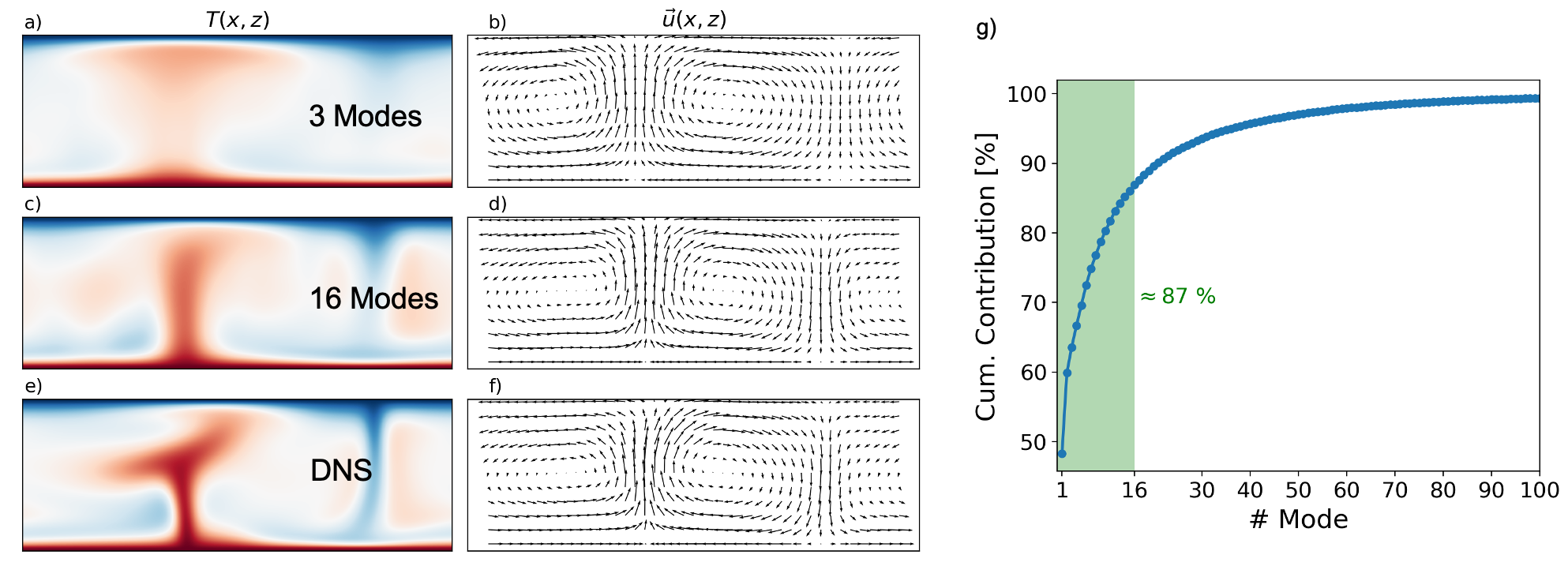}
\caption{Snapshot of two-dimensional convection flow. Panel (a,c,e) show the temperature field $T({\bm x},t_0)$, panels (b,d,f) the corresponding velocity vector field ${\bm u}({\bm x},t_0)=(u_x(x,z,t_0), u_z(x,z,t_0))$. Panels (a,b) display the reconstruction by means of the first 3 POD modes, (c,d) of the first 16 POD modes, and (e,f) display the full snapshot. (g) Cumulative spectrum of the Proper Orthogonal Decomposition. The shaded region indicates the chosen cutoff mode at $N_{\rm dof} = 16$. The first three POD modes include 63\% of the total energy, the first 16 modes include 87\%.}
\label{fig:pod}
\end{figure*}

\section{Turbulent flow}
\subsection{Model equations and parameters}
We consider a two-dimensional Rayleigh-B\'enard system where a fluid is enclosed between two impermeable plates with constant temperature difference $\Delta T=T_{\rm bot}-T_{\rm top}>0$ \cite{Chilla2012}. The Boussinesq equations connect the incompressible velocity field ${\bm u}=(u_x,u_z)$ with the temperature and pressure fields, $T$ and $\mathcal{P}$. They are given by
\begin{align}
   \label{eq:gov_eq1}
   {\bm \nabla}\cdot {\bm u} &= 0,\\
    \label{eq:gov_eq2}
   \frac{\partial {\bm u}}{\partial t} +({\bm u}\cdot {\bm \nabla}) {\bm u} &= -{\bm \nabla}P+
   \nu\, {\bm \nabla}^2{\bm u} + g\alpha (T-T_0) {\bm e}_z,\\
   \label{eq:gov_eq4}
   \frac{\partial T}{\partial t} + ({\bm u}\cdot {\bm \nabla}) T &= \kappa {\bm \nabla}^2 T\,.
\end{align}
Here, $\alpha$, $g$, $\nu$, and $\kappa$ are the thermal expansion coefficient, the acceleration due to gravity, the kinematic viscosity, and the thermal diffusivity, respectively. We set $P=\mathcal{P}/\rho_0$ in eq. \eqref{eq:gov_eq2}. Furthermore, $T_0$ and $\rho_0$ are {\em constant} reference values of the temperature and mass density, respectively. These equations stand for the differential balances of mass density \eqref{eq:gov_eq1}, momentum density \eqref{eq:gov_eq2}, and energy density \eqref{eq:gov_eq4} of a fluid parcel. In the Boussinesq approximation, it is assumed that the fluid is incompressible (or divergence-free) and that small deviations of the density of the fluid from the reference values depend linearly on temperature deviations \cite{Chilla2012}. This leads to the last term in the momentum equation \eqref{eq:gov_eq2} which couples the temperature field with the vertical velocity component.

The system is made dimensionless by the choice of the free-fall velocity scale $U_f = \sqrt{\alpha g \Delta T H}$, the free-fall time scale $T_f=H/U_f$, and $\Delta T$. Here, $H$ is the height of the convection layer, the characteristic spatial scale in the setting. In this way all material parameters and scales can be summarized in dimensionless parameters that determine the operating point of the turbulent flow. These parameters are the Rayleigh and the Prandtl numbers, ${\rm Ra} = \alpha g \Delta T H^3/(\nu \kappa)$ and ${\rm Pr} = \nu/\kappa$. They take values of ${\rm Ra}=10^5$ and ${\rm Pr}=10$ in the present proof-of-concept study. Alternatively, one can use $U_{\rm diff}=\kappa/H$ as a characteristic velocity \cite{Lorenz1963,Pfeffer2022}. This does not affect the physical outcome. Figure \ref{fig:RBC} illustrates the configuration which we want to investigate in the following by the hybrid quantum-classical algorithm.

We conduct direct numerical simulations using the spectral element solver Nek5000 \cite{nek5000} to solve the Rayleigh-B\'{e}nard system \eqref{eq:gov_eq1}--\eqref{eq:gov_eq4} in a domain $A= L \times H$ with aspect ratio $\Gamma = L/H= 2\sqrt{2}$. Dimensionless coordinates are thus $x\in [0, 2\sqrt{2}]$ and $z\in [0,1]$. Dirichlet boundary conditions are imposed for the temperature field at top and bottom, $T(z=0)=1$ and $T(z=1)=0$. Furthermore, we choose free-slip boundary conditions for the velocity field in $z$-direction, $u_z(z=0,1)=0$ and $\partial u_x/\partial z=0$ at $z=0, 1$. Periodic boundaries for all fields are taken in the horizontal $x$--direction. The chosen boundary conditions, aspect ratio and Prandtl number correspond to a popular standard case for the Lorenz systems \cite{Pfeffer2022}, but at a higher $Ra$ and thus fully turbulent in contrast to our previous work. 

\subsection{Turbulent fluctuations and heat transfer}
A central physical question in turbulent convection flows is that on the turbulent transfer of heat and momentum across the convection layer in dependence on the two dimensionless parameters, the Rayleigh number ${\rm Ra}$ and the Prandtl number ${\rm Pr}$. In response to both, these transfers can be summarized in two further dimensionless parameters, the Nusselt number ${\rm Nu}$ for the turbulent heat transfer and the Reynolds number ${\rm Re}$ for the turbulent momentum transfer. They are given by
\begin{equation}
{\rm Nu}=\dfrac{\langle u_z T\rangle_{x,t} - \kappa \dfrac{\partial \langle T\rangle_{x,t}}{\partial z}}{\dfrac{\kappa \Delta T}{H}}\;\;\mbox{and}\;\;
{\rm Re}=\frac{u_{\rm rms} H}{\nu}\,.
\end{equation}
The symbol $\langle\cdot\rangle_{x,t}$ stands for a combined average with respect to horizontal $x$-direction with $x\in [0,\Gamma]$ and time $t$. The root mean square (rms) velocity is given by $u_{\rm rms}=(\langle u_x^2+u_z^2\rangle_{A,t})^{1/2}$ where the average is now a combination of averages with respect to the simulation plane $A$ and time $t$. The two terms in the definition of the Nusselt number stand for two heat currents across the layer, the convective and the diffusive one. Their sum is constant and equal to ${\rm Nu}$ for each height $z\in [0,1]$. However their contributions to the total turbulent heat flux differ with respect to $z$, caused for example by the boundary condition $u_z=0$ at the bottom and top.

It is exactly the mean profile of the convective heat flux $\langle u_z T\rangle_{x,t}$ as a function of $z$, which we want to obtain as a reservoir computing model output. The same holds for the vertical profiles of the root mean square velocity and temperature. This is the low-order statistics of the turbulent flow which is generated by the different reservoir computing models without a knowledge of the nonlinear Boussinesq equations of motion. We will come back to these results in Sec. V D.

\section{Data reduction to latent space}
The numerical simulations are performed on a non-uniform grid of size of $384 \times 96$ points with a 2nd-order equidistant time stepping of $\Delta t= 5 \cdot 10^{-4}$.  For the analysis, the simulation data were interpolated to a uniform grid of size $N_x\times N_z = 128 \times 32$. The data set consists of a sequence of snapshots of the fields ${\bm u}(x,z,t_m)$ and $T(x,z,t_m)$ with $m=1,2, ...,10^4$; they are equidistant in time with $\tau=0.25 H/U_f$. The sequence covers the statistically stationary regime of the turbulent convection flow.  The three turbulent fields possess an input vector of size $12288$. In order to circumvent high computational effort for the machine learning, we add a pre-processing step. 

Here, we apply a POD in the form of a snapshot method \citep{Sirovich1987,Bailon2011,Pandey2020}. It is a linear method, where the data reduction is realized by a truncation to a set of Galerkin modes.  For this, we decompose the physical fields into time mean and fluctuations,  
\begin{align}
u_x(x,z,t)  &= \Tmean{u_x}(x,z) + u_x^{\prime}(x,z,t)\,,\\ 
u_z(x,z,t)  &= \Tmean{u_z}(x,z) + u_z^{\prime}(x,z,t)\,,\\ 
T(x,z,t) &= \Tmean{T}(x,z) + \theta^{\prime}(x,z,t)\,.
\end{align}
Finally, we perform the snapshot POD to the fluctuating component fields $g_k$ into time dependent coefficients $a_i(t)$ and spatial modes $\Phi_i^{\rm (k)}(x,z)$, such that the truncation error is minimized. The degrees of freedom $\tilde{N}_{\rm dof}$ can then be reduced, by taking only $\tilde{N}_{\rm dof} \ll N_{\rm dof}$ modes and coefficients with the most variance into account,
\begin{equation}
    \label{eq:pod_rom}
    g_k(x,z,t) \approx \sum\limits_{i=1}^{N_{\rm dof}} a_i(t) \Phi_i^{\rm (k)}(x,z)\,,    
\end{equation}
with $g_k=\{u_x^{\prime}, u_z^{\prime}, \theta^{\prime}\}$ and thus $k=1, 2, 3$. Figure \ref{fig:pod}(a--f) compares the reconstruction of the temperature and velocity fields from 3 and 16 modes with the original simulation data at a time instant. The time series of the expansion coefficients of three first POD modes, $a_1(t)$ to $a_3(t)$, will be fed into the recurrent network in the reconstruction phase after training. 

In the following, we use the more compact notation ${\bm a}^t=(a_1(t), a_2(t), ..., a_M(t))$ with $M=N_{\rm dof}$ and the discrete snapshot time superscript $t$. This is our dynamical system state which has to be learned by the reservoir computing model. We choose the cutoff at $N_{\rm dof}  = 16$ and capture $87\%$ of the variance of the original fields as seen in Fig. \ref{fig:pod}(g). It also implies that the notation $\rho(t)$ changes to $\rho^t$ and so on, see again eq. \eqref{evolution}.

\begin{figure*}
\includegraphics[scale=0.5]{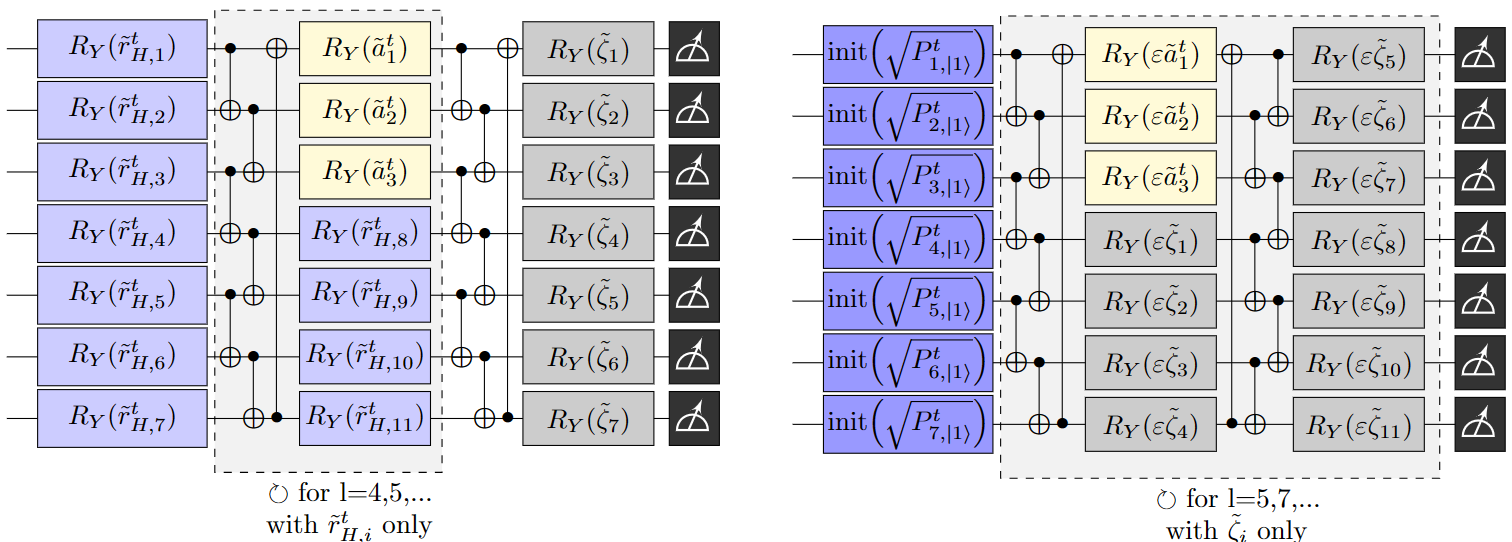}
\caption{Quantum circuits of the two applied hybrid quantum classical algorithms which represent the quantum computing version of the reservoir.  Left: hybrid quantum-classical reservoir computing algorithm H1, which represents our previous approach of ref. \cite{Pfeffer2022} with a new hyperparameter $l$ to denote the circuit complexity (or depth). Right: hybrid quantum-classical reservoir computing algorithm H2 which we will in short denote to as the full quantum reservoir computing model. The new structure incorporates an efficient pseudo-initialization of the last reservoir time step as well as an efficient memory implementation by the leaking rate $\varepsilon$, which now scales the dynamical activation of the approach without classical post-processing. In both cases, every qubit of the quantum register starts in the basis state $\ket{0}$ to the left. Each quantum circuit is shown for 7 qubits and $l=3$ layers of $R_Y$ rotation gate at a discrete time step $t$. The CNOT gates between the $R_Y$ gate connect neighboring qubits and generate entanglement. Since this pairwise 2-qubit operation is done successively over all qubits, one ends up with a fully entangled 7-qubit quantum state.  The tilde symbol indicates that the rotation gate arguments are re-scaled to cover the respective range, as given in the text. Also, in the right circuit diagram $\mbox{init}(\omega)=R_Y(2\arccos(\omega))$. The last column of symbols (to the right) in both panels stands for the measurement to read out the reservoir state.}
\label{fig:algorithm}
\end{figure*}

\section{Reservoir computing frameworks}
A reservoir computing model \cite{Jaeger2001,Jaeger2004} is one realization of recurrent machine learning beside long short-term memory networks or gated recurrent units \cite{Hochreiter1997,Chung2014}. Its fundamental element is the reservoir state vector ${\bm r}^t \in \mathbb{R}^{N_{res}}$, which evolves from $t$ to $t+1$ by a certain update equation characterizing the approach. As indicated in the introduction, this work compares three RCMs, algorithm C a classical RCM \cite{Pandey2020,Pandey2020a,Heyder2021,Pandey2022,Valori2022,Heyder2022} with linear memory and nonlinear activation, see eq. \eqref{eq:c_rcm}, algorithm H1 a hybrid quantum-classical RCM \cite{Pfeffer2022} with classical linear memory and nonlinear quantum dynamics, see eq. \eqref{eq:h_rcm}, and algorithm H2 a full quantum RCM which induces memory by reduced gate parameterization, see eq. \eqref{eq:q_rcm}. Figure \ref{fig:algorithm} illustrates the different quantum circuits of H1 (left) and H2 (right). All algorithms run here in the open-loop or reconstruction mode \cite{Lukosevicius2021}, where the reservoir receives ${\bm a}_{\rm in}^t = (a_1^t,a_2^t,a_3^t)$ at each time step and outputs the missing expansion coefficients at the next time step, $\hat{\bm a}_{\rm out}^{t+1}=(\hat{a}_4^{t+1}\,\dots,\hat{a}_{N_{\rm dof}}^{t+1})$ with $N_{\rm dof}=16$ for our study. The hat symbol identifies always the network prediction in the following. Note that in this scenario the reservoir receives the coefficients corresponding to the large-scale fluid motion while returning the small-scale features of the higher modes.

\subsection{Classical reservoir computing algorithm} 
The classical algorithm C is characterized by the following iterative equation for the reservoir state vector ${\bm r}_C$ \cite{Pathak2018,Pandey2020,Heyder2022},
\begin{equation}
	{\bm r}_{\rm C}^{t+1}= (1-\varepsilon){\bm r}_{\rm C}^t + \varepsilon\tanh\left[W^{\rm in}{\bm a}_{\rm in}^t + W^{\rm res} {\bm r}_{\rm C}^{t}\right].
 \label{eq:c_rcm}
\end{equation}
where $W^{{\rm res}} \in \mathbb{R}^{N_{\rm res} \times N_{\rm res}}$ is the sparsely occupied random reservoir matrix and $W^{\rm in}\in \mathbb{R}^{N_{\rm res}\times N_{\rm in}}$ is the random input matrix. The scalar $\varepsilon \in [0,1]$ is called leaking rate, as it scales the influence of the first memory term and the second dynamical term. This discrete time stepping from snapshot $t$ to $t+1$ comprises external forcing by the inputs ${\bm a}_{\rm in}^t$ as well as a self-interaction with the reservoir state ${\bm r}_{\rm C}^{t}$. The hyperbolic tangent $\tanh\left(\cdot\right)$ is the nonlinear activation function, which is applied to the elements of its argument vector. The reservoir matrix $W^{\rm res}$ is specified further by the reservoir density $D$, its sparsity, and the spectral radius $\varrho(W^{\rm res})$, the magnitude of the maximum eigenvalue. The hyperparameters of the classical RCM are thereby $N_{\rm res}$, $\varepsilon$, $D$ and $\varrho(W^{\rm res})$. In all reservoir computing models, the optimization is done directly for the output layer only which is represented by the matrix $W^{\rm out \ast}\in \mathbb{R}^{N_{\rm out} \times N_{\rm res}}$. The asterisk stands for the optimized matrix after training, see refs. \cite{Pandey2020,Pfeffer2022} for the details. Also there, the Tikhonov parameter $\gamma$ is explained further, which is used as a prefactor of an additional penalty term in the cost function. This hyperparameter is only applied in the classical RCM.

To give an explicit example for the one-step reconstruction mode in which the model will be used here: the classical RCM utilizes eq. \eqref{eq:c_rcm} to calculate ${\bm r}_{\rm C}^{t+1}$ from ${\bm r}_{\rm C}^t$ and ${\bm a}_{\rm in}^t$ with its three components to obtain the estimate
\begin{equation}
\hat{\bm a}_{\rm out}^{t+1} = W^{\rm out \ast} {\bm r}_{\rm C}^{t+1} \,,
\label{eq:crcm}
\end{equation}
with $\hat{\bm a}_{\rm out}^{t+1}=(\hat{a}_4^{t+1},\dots,\hat{a}_{16}^{t+1})$ in the present turbulent convection flow study.

\subsection{Hybrid quantum-classical reservoir computing algorithm 1} 
The algorithm H1 follows the classical reservoir iteration procedure  \eqref{eq:c_rcm} closely, but the update equation receives another dynamical part. We introduced H1 in \cite{Pfeffer2022}. It is given by
\begin{subequations}
\label{eq:h_rcm}
\begin{equation}
{\bm r}_{\rm H1}^{t+1}= (1-\varepsilon){\bm r}_{\rm H1}^t + \varepsilon {\bm p}^{t+1} \,,
\end{equation}   
with 
\begin{equation}
p_j^{t+1}=\rho_{jj}^{t+1}=\big| \langle e_j|U({\bm \zeta},{\bm a}_{\rm in}^t,{\bm r}_{\rm H1}^t)|e_1\rangle\big|^2\,,
\end{equation}
\label{eq:h_rcm}
\end{subequations}
where $|e_j\rangle$ is the $j$-th basis vector of the $n$-qubit Hilbert space $\mathbb{C}^{\otimes n}$ with a dimension $N=2^n$. Furthermore, $|e_1\rangle=|0\rangle^{\otimes n}$ is the $n$-qubit basis vector for which every of the $n$ individual qubits is in the base state $|0\rangle$. The quantum circuit $U({\bm \zeta},{\bm a}_{\rm in}^t,{\bm r}_{\rm H1}^t)$ is parameterized with random values ${\bm \zeta}$, the current input ${\bm a_{\rm in}^t}$ and the last reservoir state ${\bm r}_{\rm H1}^t$. We initialize ${\bm r}_{\rm H1}^0$ as a uniform-amplitude probability vector at all entries, which guarantees that ${\bm r}_{\rm H1}^t$ will remain a probability vector for all times $t$. 

The structure of the circuit is a repeated pattern of $R_Y$-gate and separating CNOT-gate layers, where the arguments of the $R_Y$-gates are the only difference between the layers. The circuit always starts and ends with an $R_Y$-gate layer, as shown in Fig. \ref{fig:algorithm} (left). As a new hyperparameter, we introduce the depth $l$ as the amount of $R_Y$-gate layers inside the circuit. An $R_Y$-gate layer applies an $R_Y$-gate on every qubit, where the arguments are scaled versions of the aforementioned variables. The random values $\bm \zeta$ vary in $[0,4\pi]$, the probabilities ${\bm r}_{\rm H1}^t$ are multiplied with $\pi$ to vary in $[0,\pi]$ and the inputs ${\bm a}_{\rm in}^t$ are re-scaled to vary in $[-\pi,\pi]$. For example, the random values $\bm \zeta$ can  build various unitary matrices, as the matrix definition of the $R_Y(\varphi)$ gate is given by
\begin{equation}
\label{Ry}
R_Y(\varphi) = \begin{pmatrix}
    \cos(\varphi/2) & -\sin(\varphi/2) \\
    \sin(\varphi/2) & \cos(\varphi/2)
\end{pmatrix}.
\end{equation}
Note that the loading of the reservoir state and the dynamical system modes by rotations into H1 and H2 introduces a nonlinearity in the (linear) quantum dynamics. 

The CNOT-gate layer uses $n$ CNOT gates such that every qubit is control and target qubit once. The specific arrangement is random but fixed. We always assure that it is impossible to separate any subgroup of qubits from the remaining one; see \cite{Pfeffer2022} on the importance of the entanglement for the reconstruction quality measured by the mean squared error. The sorting  of the $R_Y$ gate arguments is such that the last layer is filled with random values, all other layers receive probabilities, and the first three entries in the second layer are always the inputs. There are too many possibilities to proof that this specific circuit construction is the optimum, but we tested many architectures and choose the described due to best overall performance.

\subsection{Hybrid quantum-classical reservoir computing algorithm 2}
The new hybrid algorithm H2 is modified such that the complete execution on a quantum computer is enabled. It is given by 
\begin{subequations}
\label{eq:q_rcm}
\begin{equation}
    {\bm r}_{\rm H2}^{t+1}=  \bm p^{t+1}
\end{equation}
with
\begin{equation}
p_j^{t+1}=\rho_{jj}^{t+1}=\big| \langle e_j|U(\varepsilon \bm \zeta,\varepsilon \bm a_{\rm in}^t)|\bm{\hat{r}}_{\rm H2}^t\rangle\big|^2\,. 
\end{equation}
\end{subequations}
An identity transformation is imposed by
\begin{equation}
U(0,0)=\mathcal{I}_N\,.
\label{eq:q_rcm1}
\end{equation}
This is a consequence of the inclusion of the leaking rate $\varepsilon$ in the arguments for H2, the major difference to H1. Furthermore, the approximate reservoir vector is
\begin{equation}
\ket{\bm{\hat{r}}_{\rm H2}^t} = \bigotimes_{i=1}^n  \left[
        \sqrt{1-P^t_{i,\ket{1}}} |0\rangle + 
         \sqrt{P^t_{i,\ket{1}}} |1\rangle \right]\,,
\label{eq:q_rcm2}
\end{equation}
where $P^t_{i,\ket{1}}$ denotes the probability of measuring qubit $i$ of the whole $n$-qubit register in basis state $\ket{1}$ at reservoir time step $t$. In other words, the two probabilities, $1-P^t_{i,\ket{1}}$ and $P^t_{i,\ket{1}}$ are the diagonal elements of the $2\times 2$ density matrix $\tilde{\rho}_i^t$ which is obtained by the following partial trace of the original density matrix of the $n$-qubit state, $\rho^t$, which traces out all qubits except qubit $i$. It is given by
\begin{equation}
\tilde{\rho}_i^t={\rm Tr}_{1,2,\dots,i-1,i+1,\dots,n}(\rho^t) \,.
\label{eq:q_rcm3}
\end{equation}
In a quantum algorithm, this is realized by an individual measurement at qubit $i$ of the whole quantum register only. We thus structure the circuit such that it is initialized by a completely separable approximation of the last reservoir state and thus ease the reservoir initialization at each step, see refs. \cite{Mujal2023,Cindrak2023} for alternative solutions to circumvent this bottleneck in analogue quantum reservoir computing. This combines two advantages: It is a minimal initialization for the quantum circuit with only $n$ operations which contains the integral information on the reservoir.  Combined with the adapted definition of the unitary $U$ in eq. \eqref{eq:q_rcm}, we include the previously external memory in the quantum circuit iteration step. That is, the dynamics will correspond to an identity transformation, see eq. \eqref{eq:q_rcm1}, once the leaking rate is $\varepsilon=0$.

Here, the structure of the quantum circuit starts with the preparation of $\ket{\bm \hat{r}_{\rm H2}^t}$, which comes down to an $R_Y$-gate layer parameterized with the $P_{i,\ket{1}}^t$, as done in upcoming eq. \eqref{Eq:R2}. The following circuit layer has pairwise CNOT-gate layers, where every second CNOT layer is the inverse of the first CNOT layer, as indicated in the right panel of Fig. \ref{fig:algorithm}. Thereby, we satisfy the central condition of eq. \eqref{eq:q_rcm1}, as $R_Y(0)=\mathcal{I}_2$. Beside the input gates, all $R_Y$ gates are filled with random values $\zeta_k$. Further details of the circuit are adopted from H1. Note however, that the circuit ends with a CNOT layer for an even number of $R_Y$ layers.

\subsection{Advantages of hybrid reservoir computing approaches} 
There are several aspects which motivate the presented  hybrid quantum-classical approaches H1 and particularly H2. First, we want to investigate whether the nonlinear activation of the quantum circuits is preferable over $\tanh(\cdot)$ of eq. \eqref{eq:c_rcm}. The inherent nonlinearity of product concatenations from the unitary gates to larger matrices approximates the inherent nonlinearity of the original flow problem, see the second term on the left hand side of eq. \eqref{eq:gov_eq2}.

Secondly, we note that for online processing of the data, we can strongly reduce the required memory once the output of a time step is generated. For instance, one needs to save the used probabilities only, i.e., for H1 the $r_{H,i}^t$ and for H2 the $P_{i,\ket{1}}^t$, to recompute the reservoir state at the subsequent time step again. This would not be possible for the classical reservoir. Especially for H2, this always means that for a reservoir of size $N$, one keeps $\log_2(N)$ values only to store sufficient information of the reservoir evolution. If necessary, it is then possible to reconstruct all $2^n$ reservoir states in parallel.

Thirdly, H2 is in contrast to H1 fully executable on a quantum computer, as the leakage rate of the reservoir step is included in the rotation gate parameters of the quantum circuit. Therefore, no external memory is required, cf. eq. (\ref{eq:h_rcm}a) for H1. Additionally, it might be possible to extend the present scheme H2 to a multi-stepping approach on the quantum computer if one implements the output weights as well. This could be done by encoding the output on an ancilla qubit to either measure it directly or insert it back for the next time step via a rotation controlled by this ancilla qubit. Future work in this direction might elaborate whether this approach can strongly reduce the computational complexity in comparison to the classical approach by reducing the number of sampling steps (or shots) for multiple time steps in the hybrid case.

\section{Comparison of the models}
\subsection{Error quantification measures}
\begin{figure*}
\centering
\includegraphics[scale=0.62]{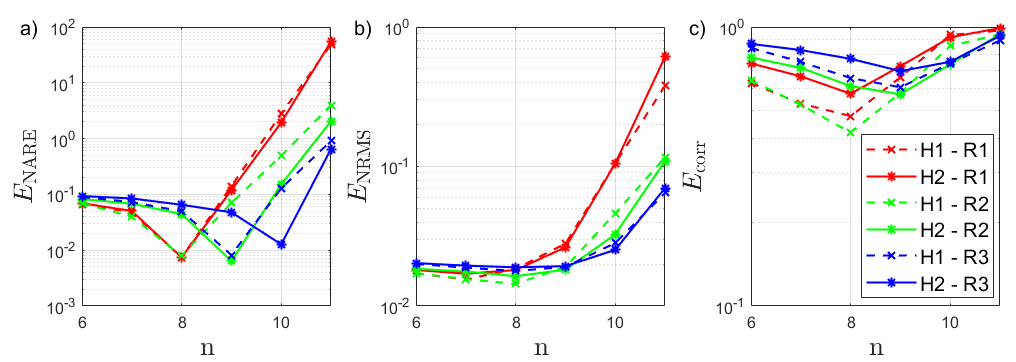}
\caption{Comparison of three error measures, which are defined in eqns. \eqref{eq:SNMSE}--\eqref{eq:E_nare} and three amplitude encoding methods, which are defined in Eqns. \eqref{Eq:R1} to \eqref{Eq:R3} for both hybrid RCM cases, H1 and H2. They are termed $R_1$, $R_2$, and $R_3$. Here, each data point is the median obtained from ten different random reservoir seeds ${\bm \zeta}$. The hyperparameters $l$ and $\varepsilon$ are optimal in each case. For the optimization, $2\le l \le 7$ and $0.05\le \varepsilon \le 1$. We trained for 5000 and evaluated on the subsequent 500 discrete time steps.}
\label{fig:ErrorMeasures}
\end{figure*}
In order to evaluate the reconstruction quality of the proposed RCMs, we need appropriate error measures. The first standard is the root mean squared error of the prediction. Since multiple modes have to be reconstructed, we take the normalized error with respect to each mode and combine the $N_{\rm out}=13$ individual errors of the reconstructed modes to the normalized root mean squared error (NRMS). This results to
\begin{equation}
    E_{\rm NRMS} = \frac{1}{N_{\rm out} {\cal T}} \sum\limits_{j=N_{\rm in}+1}^{N_{\rm dof}} \, \left[\frac{ \sqrt{\sum\limits_{t=t_a}^{t_b} \left( a_j^t-\hat{a}_j^t\right)^2}}{ 
\underset{t}{\text{max}}(a_j^t)-
\underset{t}{\text{min}}(a_j^t)} \right]   
\label{eq:SNMSE}
\end{equation}
where ${\cal T}=t_b-t_a$ is the length of the testing phase (measured in discrete time steps as discussed in Sec. III). The maxima and minima are determined with respect to the time interval $[t_a,t_b]$. 

A second popular approach is the correlation error, also known as the coefficient of determination, or R2--score, which computes the correlation between the original and predicted modes \cite{Mujal2023}. Here, we average the square of correlation, such that the correlation error is given by
\label{eq:E_corr}
\begin{equation}
\label{eq:E_corr}
    E_{\rm corr} = 1 - \frac{1}{N_{\rm out}} \sum\limits_{j=N_{\rm in}+1}^{N_{\rm dof}} \left( \frac{\text{cov}(a_j^t,\hat{a}_j^t)}{\sigma(a_j^t) \sigma(\hat{a}_j^t)} \right)^2
\end{equation} 
Here, $\sigma(\cdot)$ is the standard deviation and cov$(\cdot,\cdot)$ the covariance of the arguments. Note that by the square of the correlation, we value anti-correlation as much as positive correlation, thus $0\le E_{\rm corr}\le 1$. Strongly correlated time series send $E_{\rm corr}\to 0$. Both error measures are applicable to any dynamical system. However in the present work, we consider a turbulent flow; the RCM application is focused to reconstructed statistical properties, as motivated in the introduction and in Sec. II. Such properties can be the mean vertical profiles of the velocity components or the temperature field. 

Therefore, we define an additional measure which is directly related to the low-order statistical reconstruction results, the normalized average relative error (NARE), which has been used in classical RCM applications \cite{Srinivasan2019,Pandey2020} and is given by the L$_1$ norm, 
\begin{equation}
     E_{\rm NARE} = \frac{1}{C_{\rm max}}\int\limits_0^1 \big|\langle u_z T \rangle_{x,t}-\langle \hat{u}_z \hat{T} \rangle_{x,t}\big| \, dz
    \label{eq:E_nare}
\end{equation}
with the normalization constant 
\begin{equation}
C_{\rm max}=2 \max_{z\in [0,1]} |\langle u_z T\rangle_{x,t}|\,.
\end{equation}
Here, $\hat{u}_z$ and $\hat{T}$ are the reconstructed flow values which are obtained by the Proper Orthogonal Expansion (POE) from the modes $\hat{a}_j^t$. As the convective heat flux is prone to the error of two fields, it is a suited measure of the accordance of the inferred convection flow. We compare these three errors for the hybrid quantum RCMs H1 and H2 in Fig. \ref{fig:ErrorMeasures} as a function of the reservoir size controlled by the qubit number $n$ and for different amplitude encoding methods, which will be detailed in the next subsection. 

First, it can be observed that $E_{\rm corr}$ is relatively large with a minimum for 8 to 9 qubits. The reason is that the accurate reconstruction of the POD modes is difficult as frequent deviations of the time series $\hat{a}_i^t$ from the ground truth are inevitable in this higher-dimensional, turbulent flow problem. In contrast to low-dimensional dynamical systems, such as the Lorenz model, a reservoir computing model will not reconstruct the exact systems trajectory in the high-dimensional phase space with a sampling time step of $0.25 t_f$, but generate a trajectory which gives the right low-order statistics. This is even the case for the classical RCMs \cite{Srinivasan2019,Pandey2020,Heyder2021,Heyder2022}. Thus it is not appropriate to optimize the network on the basis of $E_{\rm corr}$. Particularly weak correlations are not directly linked to the dynamical quality of the reconstruction. 

Secondly, we observe from the figure that both, $E_{\rm NRMS}$ and $E_{\rm NARE}$, grow eventually with a large qubit number $n$. However, it can be observed that particularly for 8, 9, and 10 qubits, the physical error $E_{\rm NARE}$ improves by almost one order of magnitude while $E_{\rm NRMS}$ remains relatively constant and insensitive. We thus conclude that while the accuracy of the reconstructed modes is difficult to improve further, the physical properties of the flow are more sensitive. Thereby, for most of the remaining analysis, we will continue our RCM evaluation with $E_{\rm NARE}$ only, as it is the physically relevant measure for the fluid mechanical application of the quantum algorithm.

\subsection{Different amplitude encodings of classical data}
Besides the hyperparameters, which will be discussed further below, the quantum circuits need to encode the classical input data $a_j^t$. This can be realized in different ways. We discuss the following three amplitude encoding methods, 
\begin{align}
&R_1:=R_Y(2\cos^{-1}(\tilde{a}_j^t))=\begin{pmatrix}
        \tilde{a}_j^t & -\sqrt{1-\tilde{a}_j^{t\,2}} \\
        \sqrt{1-\tilde{a}_j^{t\,2}} & \tilde{a}_j^t
     \end{pmatrix} 
\label{Eq:R1}\\
&R_2:=R_Y(2\cos^{-1}(\sqrt{\tilde{a}_j^t}))=\begin{pmatrix}
        \sqrt{\tilde{a}_j^t} & -\sqrt{1-\tilde{a}_j^t} \\
        \sqrt{1-\tilde{a}_j^t} & \sqrt{\tilde{a}_j^t}
     \end{pmatrix} 
\label{Eq:R2}\\
&R_3:=R_Y(2\pi\tilde{a}_j^t)=\begin{pmatrix}
        \cos(\pi \tilde{a}_j^t) & -\sin(\pi \tilde{a}_j^t) \\ \\
        \sin(\pi \tilde{a}_j^t) & \cos(\pi \tilde{a}_j^t)
    \end{pmatrix}\,.
\label{Eq:R3}
\end{align}
The tilde symbol in the equations indicates again, that the input mode $a_j^t$ needs to be re-scaled such that it only varies in the interval $[-1,1]$. Encoding $R_1$ ensures that $\tilde{a}_j^t$ is the component of the corresponding qubit, while encoding $R_2$ reveals $\tilde{a}_j^t$ after a measurement. Encoding $R_3$ is a natural encoding inside the $R_Y$-gate, where we only re-scale the input to harness the largest but still unique range of the trigonometric functions. 

Each approach induces a specific nonlinear characteristic and the superiority of each encoding may change for different learning tasks. We come back to Fig. \ref{fig:ErrorMeasures} where we plotted all three error measures versus qubit number $n$ for $R_1$ to $R_3$. The error measure $E_{\rm NRMS}$ shows an approximate independence on the specific encoding scheme for $n\le 9$ for both H1 and H2. Only for the larger qubit numbers $R_3$ performs best. In case of $E_{\rm NARE}$ a local minimum can be observed for each encoding scheme. All three amplitude encodings have their optimum at a magnitude of approximately $10^{-2}$. This is obtained for $R_1$ and $R_2$ already at a smaller number of 8 qubits. Again, for the lower reservoir dimensions of $n=6$ and $n=7$ all errors are of the same order of magnitude. As already said, for the largest reservoir dimensions, the error increases, similar to $E_{\rm NRMS}$, for $n\ge 10$. In conclusion, we do not fix the specific amplitude encoding for the following analysis, but use it as a further degree of freedom to be optimized. 

\begin{table}
\begin{center}
\begin{tabular}{lll}
\hline\hline
Algorithm & Hyperparameter & Symbol  \\ 
\hline
Classical & Reservoir density & $D$ \\
          & Spectral radius   & $\rho(W^{\rm res})$  \\ 
          & Tikhonov parameter $\;\;$ & $\gamma$  \\
Quantum (H1,H2) $\;\;\;$   & Circuit layer depth       & $l$ \\
                        & Input encoding & $R_i$ \\
Joint     & Reservoir size    & $N_{\rm res}$ \\
          & Leaking rate      & $\varepsilon$ \\
\hline\hline          
\end{tabular}
\end{center}
\caption{
\label{tab:hyp} 
List of hyperparameters for the classical and both hybrid quantum-classical algorithms. The number of used qubits follows by $n=\log_2(N_{\rm res})$ in H1 and H2. The hyperparameters, which are not plotted, in the figures are always pre-optimized, that is, we take the optimal values such that the single-best representation is illustrated.}
\end{table}
\begin{figure}
    \centering  
    \includegraphics[scale=0.6]{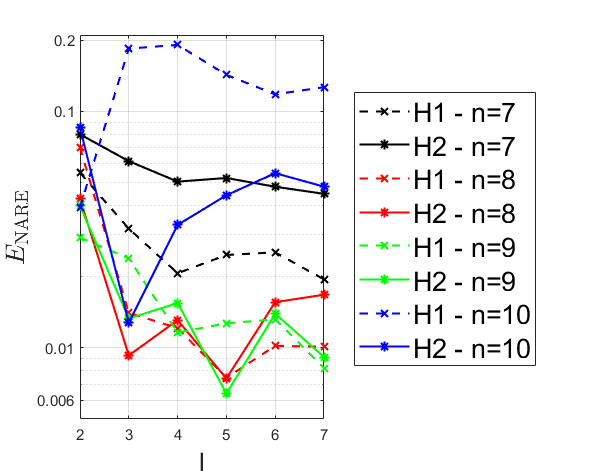}
    \caption{Evaluation of the NARE as a function of circuit depth $l$ for the hybrid quantum cases H1 and H2. Each data point is the median of $E_{\rm NARE}$, obtained from ten random reservoir seeds, using the optimal leaking rate $\varepsilon$ and optimal input amplitude encoding. In nearly all cases, the optimal values are $\varepsilon=0.2$ and $R_1$ for the input encoding.}  
    \label{fig:Hyper_Q}
\end{figure}
\begin{figure*}
    \includegraphics[scale=0.55]{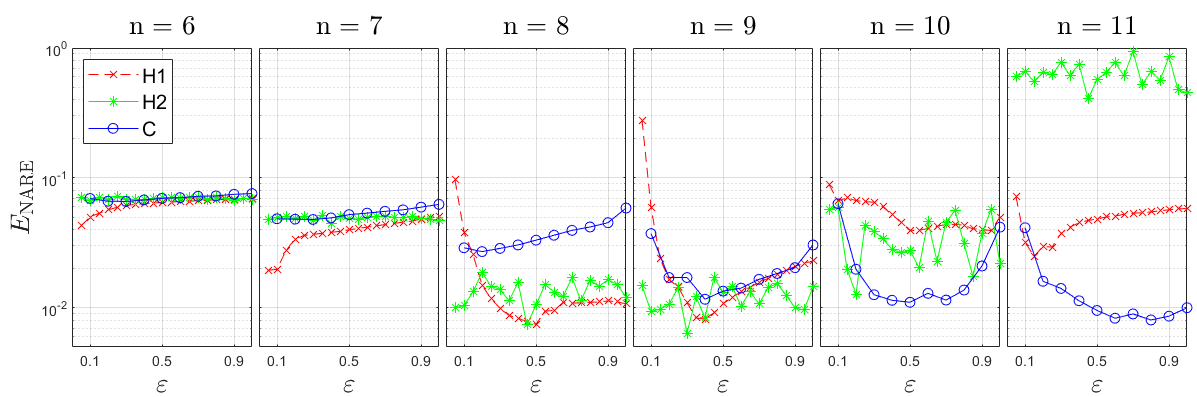}
    \caption{Comparison of the error $E_{\rm NARE}$ for the convective heat flux $\langle u_z T\rangle_{x,t}$ versus the leaking rate $\varepsilon$. We show the results for the three reservoir computing approaches C, H1, and H2 and for different reservoir sizes $2^n$. Displayed are the median values for 10 seeds with H1 and H2 and 100 seeds for $C$. The optimum of each curve is the single-best median for the respective approach at the given reservoir size. Note that the best classical reservoir performance is achieved at substantially larger reservoir sizes ($n=11$) as compared to the quantum algorithms ($n=8$ for H1 and $n=9$ for H2).}
    \label{fig:Res_Size}
\end{figure*}

\subsection{Hyperparameter dependence}
Table \ref{tab:hyp} summarizes all hyperparameters that appear either in the classical or in the hybrid quantum-classical RCMs. The hyperparameters, which exist in the quantum case only, are the circuit layer depth $l$ and the type of amplitude encoding. The latter was already discussed in the past subsection. Joint parameters of the classical and quantum case, which will be studied in the following, are the reservoir size $N_{\rm res}$ and the leaking rate $\varepsilon$. We also mention at this point that all studies for H1 and H2 are conducted with the IBM Qiskit statevector simulator where the measurements are not subject to shot noise \cite{Qiskit}. 

We train all cases for 5000 output time steps and validate the trained network on the subsequent 500 output time steps.

\subsubsection{Quantum circuit layer depth}
The quantum circuits of H1 and H2 can be parameterized by the amount of $R_Y$-layers $l$ and the positions of the CNOT-gates. The later aspect showed no significant influence during our analysis, that is, the performance of both algorithms is relatively constant as long as all qubits are connected, as described in Sec. IV B. Meanwhile, the depth of the circuit is of interest, as it is directly proportional to the computational effort of the approach, i.e., the number of operations, as well as the realizability on real quantum computers. Figure \ref{fig:Hyper_Q} displays the dependence of $E_{\rm NARE}$ on the layer depth for H1 and H2. We collect the results for $n=7, 8, 9$, and 10. Except for $n=10$, the overall trend of the error $E_{\rm NARE}$ is a decrease with growing circuit depth $l$. For $n=10$, we observe a clear advantage of H2 compared to H1. For $n=8$ and $n=9$, the  cases H1 and H2 perform similarly well, almost at their optimal error measure. We finally mention that the median was taken over a rather small number of different reservoir realizations.
\begin{figure*}
\includegraphics[scale=0.5]{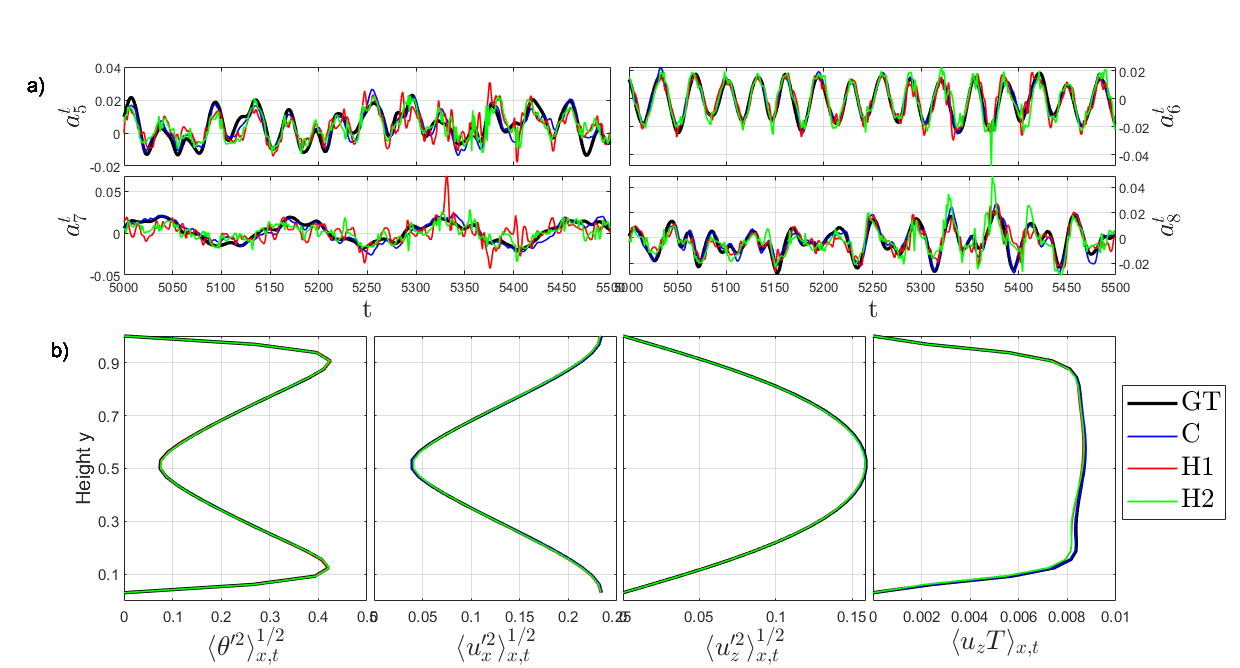}
\caption{Single-best reservoir computing results for $N_{\rm in}=3$. We compare the ground truth (GT) with a classical reservoir computing model (C) with $N_{\rm res}=2048$ and the hybrid quantum-classical quantum algorithms (H1, H2) with $N_{\rm res}=512$ and 6 layers in both cases. All networks output 500 time steps. Panels (a) display the time series of modes $a_5(t)$ to $a_8(t)$. Panels (b) show vertical mean profiles of essential low-order statistical measures. These are the  thermal variance $\langle\theta^{\prime 2}\rangle_{x,t}$, the horizontal velocity fluctuations $\langle u_x^{\prime 2}\rangle_{x,t}^{1/2}$,  the vertical velocity fluctuations $\langle u_z^{\prime 2}\rangle_{x,t}^{1/2}$, and the convective heat flux $\langle u_zT\rangle_{x,t
}$.}
\label{fig:results_10}
\end{figure*}
\begin{figure*}
\includegraphics[scale=0.45]{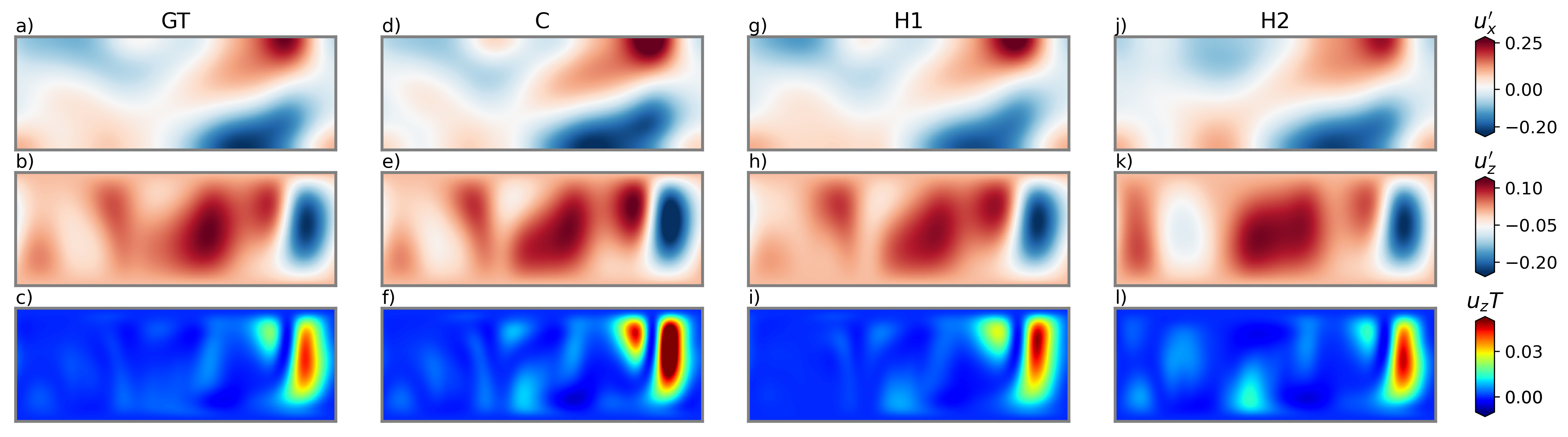}
\caption{POD reconstruction based on the latent space variables in Fig. \ref{fig:results_10}. We show the horizontal and vertical velocity fluctuation fields, $u_x'$ in panels (a,d,g,j) and $u_z'$ in panels (b,e,h,k). Furthermore, the local convective heat flux $u_zT$ is displayed in panels (c,f,i,l). The fields of the classical reservoir algorithm reconstruction are shown in panels (d-f), while the quantum algorithm results for H1 and H2 are shown in (g-i) and (j-l), respectively. For comparison, the ground truth, i.e., the POD model, is shown in panels (a-c). It can be seen that both quantum reservoirs, as well as the classical counterpart, reproduce the spatial features of all three fields of the ground truth (GT) well.}
\label{fig:results_11}
\end{figure*}

\subsubsection{Reservoir size and leaking rate}
In Fig. \ref{fig:Res_Size}, we show the median of $E_{\rm NARE}$ in dependence on the leaking rate $\varepsilon$ for different reservoir sizes $N_{\rm res}=2^n$. For all points, we averaged here over 100 seeds for C and 10 seeds for H1 and H2, while all other parameters are pre-optimized, that is, we choose the optimal spectral radius $\rho(W^{\rm res})$ and the Tikhonov parameter $\gamma$ in the classical RCM case, the optimal input encoding $R_i$ and the amount of layers $l$ in the hybrid quantum-classical cases. We choose this optimum such that the single-best median is illustrated for the respective approach and reservoir size. We observe that H1 and H2 seem to outperform the classical approach for qubit numbers  $n<10$. The global optimum, i.e., the minimal amplitudes of $E_{\rm NARE}$, are obtained for the new architecture H2 at $n=9$, though the other RCMs can perform similarly well if the reservoir is large enough.

\subsection{Statistical analysis of the reconstructed fields}
Of particular relevance in turbulent convection is the mean vertical convective heat flux profile, which is the one-point correlation of the vertical velocity component and the temperature field, $\langle u_z T (z)\rangle_{x,t}$, see Sec. II B. It is a measure of the amount of heat transported by fluid motion from the bottom of the layer to the top of the layer. Such a vertical profile is more difficult to reconstruct, as it combines the statistics of two reconstructed fields. In addition, we will monitor root mean square profiles of the velocity components and the temperature. These are essential low-order statistical properties of the flow at hand. They are also important when the turbulence cannot be modeled down to the smallest physically relevant scale and has to be parametrized. This is the case for in subgrid-scale parametrizations in global circulation models, e.g. in atmospheric turbulence \cite{Stevens2005}. 

We illustrate in Fig. \ref{fig:results_10} the single best reconstruction of each RCM model combined with the mean statistical profiles of the flow. We evaluated the previous grid search over the hyperparameters for the single best results. Figure \ref{fig:results_10}(a) displays the reconstructed time series of four POD expansion coefficients from C, H1, and H2 in comparison to the ground truth (GT). It is seen that the curves are not followed exactly, but that the overall trends and thus the low-order statistics are represented well. Note, that this is not only for the case for H1 and H2, but also for C. Illustrated are the best cases for the physical error measure $E_{\rm NARE}$. In other words, we optimized for the lower part of the figure. 

The mean convective heat flux profile in the most right panel of Fig. \ref{fig:results_10}(b) is the most sensitive, which is the reason why we utilized it for the hyperparameter optimization. This statistical correlation is connected to the hot rising and the cold falling plumes which are visible in Fig. \ref{fig:pod}(e). Figure \ref{fig:results_11} displays finally the spatial reconstruction of velocity and heat flux fields for all three cases. We find that both quantum algorithms, as well as the classical counterpart, produce statistical profiles that are in good accordance with the ground truth, for both, the root mean square profiles of the velocity components and temperature as well as for the convective heat flux. This demonstrates the applicability of the hybrid quantum-classical reservoir computing model as a reduced-order model in combination with the encoder/decoder module in the form of POD/POE, respectively.  

\section{Final Discussion}
Two central objectives can be given for the present proof-of-concept study. First, we wanted to extend the application of hybrid quantum-classical reservoir computing algorithms towards more complex classical dynamical systems. Starting with the well-known Lorenz 63 benchmark case and its extension to 8 degrees of freedom in \cite{Pfeffer2022}, we increased here the complexity of the task to be learned further by proceeding to a {\em turbulent} convection flow at the same geometry and Prandtl number as in the Lorenz cases, but at a significantly higher Rayleigh number (the latter of which measures the driving by buoyancy forces of the flow). We integrated the quantum circuit therefore into a combined encoder/decoder--reservoir computing pipeline which has to be used as well when classical machine learning is applied to turbulent flows \cite{Pandey2022}, see again Fig. \ref{fig:pipeline}. Since the phase space of the Rayleigh-B\'{e}nard system is higher-dimensional and thus the dimensionality of the turbulent attractor, the reservoir computing algorithm is only able to predict low-order turbulence statistics rather than exactly following a specific dynamical systems trajectory for a longer time. But this is exactly the task that we had in mind, reproducing 2nd-order statistics, such as the convective heat flux, in a data-driven model without solving the full nonlinear partial differential equation system of Rayleigh-B\'{e}nard convection.    

The second objective is related to the modified architecture of H2 in comparison to H1. We have compared  both hybrid quantum-classical algorithms with respect to various hyperparameters and found that they mostly perform equally well for the reconstruction tasks. Nonetheless, the update of the circuit architecture H2 can be evaluated completely on a quantum computer, which enables further steps of the hybrid algorithm on the quantum device and avoids additional external memory as H1. Additionally, the simulation with the circuit architecture H2 can be realized more efficiently than the one for H1 once the layer depth is $l>3$. The reason is that in H2 every operation that follows the input encoding can be summarized to one pre-computable matrix which acts on the time-dependent inputs. This is not possible for the architecture H1, as further subsequent circuit layers also have to be filled with the time-dependent components of the probability amplitude vector ${\bm p}^t$. We also investigated, if a random value encoding which is used in H2 works for the original hybrid algorithm H1, but it was found that this method strongly impairs the performance of H1. It can be concluded therefore that H2 is a more efficient implementation of the reduced-order model of the turbulent convection flow by means of a hybrid quantum-classical reservoir computing algorithm; it is comparable to the best-case scenarios of the classical reservoir computing approaches which however need at least twice as large reservoirs in the present case, see Fig. \ref{fig:Res_Size}. In future applications, this could reduce the numerical effort for both, hyperparameter optimization and production runs.

The demonstration of the capabilities and potential of the present framework, but also its current limitations, has to our point of view its value for following studies in this subject in the future applied to realistic fluid flow problems.

A first open point for our future work is to solve the sampling problem. We used the ideal Qiskit statevector simulator for both algorithms, H1 and H2, which circumvents the crucial problem of approximating the necessary probabilities \cite{Qiskit}. A deeper analysis of this aspect shows that the computational overhead to approximate the probabilities, both, for the best cases of H1 and H2, is big. In detail, more than $2^{20}$ samples (or shots) are necessary for comparable results. This seems to damp the prospects for an application on current noisy intermediate-scale quantum devices. However, a repetition of the hyperparameter grid search with sample-based probabilities and the additional implementation of weak measurements as in ref. \cite{Mujal2021} might ease this problem. Furthermore, it has to be evaluated if the hybrid reservoir computing approach can be further scaled up to more vigorous turbulence, i.e., flows at higher Rayleigh number. This would imply a higher dimension of the latent data space and possibly a different encoding/decoding scheme, see  \cite{Pandey2022,Heyder2022} for the classical cases. These investigations are going on and will be reported elsewhere.

\acknowledgments
This work is supported by the project no. P2018-02-001 "DeepTurb -- Deep Learning in and of Turbulence" of the Carl Zeiss Foundation and by the Deutsche Forschungsgemeinschaft under grant no. DFG-SPP 1881. We acknowledge support for the publication costs by the Open Access Publication Fund of the Technische Universität Ilmenau.

\appendix
\label{sec:materials_methods}

\section{Quantum computing basics}
In this appendix, we summarize some basic definitions of quantum computing. Further details can be found in the textbook of Nielsen and Chuang \cite{Nielsen2010} or a review by Bharadwaj and Sreenivasan \cite{Bharadwaj2020}. While a single classical bit can take two discrete values only, namely $\{0,1\}$, a single quantum bit (in short qubit) is a superposition of the two basis states of the vector (or better Hilbert) space $\mathbb{C}^2$. This is sometimes illustrated as an arbitrary point on the surface of the so-called {\em Bloch sphere} (a unit sphere). One writes
\begin{equation}
|q_1\rangle=c_1|0\rangle+c_2|1\rangle=c_1
\left(
\begin{array}{c}
1 \\ 0\\
\end{array}\right)
+c_2
\left(
\begin{array}{c}
0 \\ 1\\
\end{array}\right)\,, \label{Eq:B1}
\end{equation}
with $c_1, c_2\in \mathbb{C}$ and $\sqrt{|c_1|^2+|c_2|^2}=1$. Vectors $|0\rangle$ and $|1\rangle$ are the basis vectors in Dirac notation \cite{Nielsen2010}. A qubit can be considered as the simplest quantum system. In other words, the qubit can be consequently found in infinitely many superposition states, all the points that fill the surface of the unit sphere. It is the building block of an $n$-qubit system, also denoted as an $n$-qubit quantum register. They are formed by successive tensor products of qubits. For example, a two-qubit state vector is the tensor product of two single-qubit vectors, 
\begin{equation}
|q_1\rangle\otimes |q_1^{\prime}\rangle \in \mathbb{C}^2\otimes\mathbb{C}^2\,.
\end{equation}
The basis of this tensor product space is given by 4 vectors: $|{\bm j}_1\rangle=|0\rangle\otimes |0\rangle$, $|{\bm j}_2\rangle=|0\rangle\otimes |1\rangle$, $|{\bm j}_3\rangle=|1\rangle\otimes |0\rangle$, and $|{\bm j}_4\rangle=|1\rangle\otimes |1\rangle$. An $n$-qubit quantum state $|{\bm \Psi}\rangle $ is consequently an element of a $2^n$-dimensional tensor product Hilbert space ${\cal H}=(\mathbb{C}^2)^{\otimes n}$. The state vector is given by
\begin{equation}
   \ket{\bm \Psi} = \sum\limits_{k=1}^{2^n} c_k \ket{{\bm j}_k} \hspace*{1em} \text{with} \hspace*{1em} \sum\limits_{k=1}^{2^n} |c_k|^2 = 1\,.
\end{equation}
An $n$-qubit state vector is called {\em fully separable} if it can be written as a tensor product,
\begin{equation}
   \ket{\bm \Psi} = \overset{n}{\underset{i=1}{\bigotimes}} \,\ket{q_i} \,,
\end{equation}
where $\ket{q_i}$ are single qubit quantum states given by eq. \eqref{Eq:B1}. It is called {\em separable} if a tensor product decomposition of $\ket{\bm \Psi}$ into blocks is possible with at least one multi-qubit quantum state $\ket{q_i}$, that is not fully separable. Multi-qubit quantum states which are not separable are called {\em entangled}. An $n$-qubit quantum state is called {\em fully entangled} if no subspace of separable qubits exists. Fully entangled quantum states are characterized by correlations which do not classically exist. The entanglement is a unique property of quantum computing; it is supposed to be responsible for the quantum advantage of some quantum algorithms with respect to their classical counterparts, such as prime factorization \cite{Nielsen2010}, as single qubit operations act non-trivially on a large quantum state and produce global parallel processing by a single computational step.

The time evolution of a quantum state is described by a unitary transformation,
\begin{equation}
   \ket{\bm \Psi}(t) = U(t) \ket{\bm \Psi}(0) \quad \mbox{with}\quad U(t)^{-1}=U(t)^{\dagger}\,.
\end{equation}
Elementary unitary transformation are supplied by {\em quantum gates}, for example rotations of a single qubit. Note that they are reversible transformations. The $R_Y$ rotation gate is defined by eq. \eqref{Ry} in Sec. IV B. A second central gate is the controlled NOT gate (in short CNOT) which connects two qubits. It represents a flip of the target qubit once the control qubit is in state $|1\rangle$. The logical table of the two-qubit CNOT gate is shown in table \ref{tab:CNOT} which can be transformed into a 4x4 unitary matrix.
\begin{table}
\begin{center}
\begin{tabular}{cccc}
\hline\hline
\multicolumn{2}{c}{Input}&\multicolumn{2}{c}{Output}\\
\hline
Control & Target  & Control  & Target  \\
$\ket{q_0}$ & $\ket{q_1}$ & $\ket{q_0}$ & $\ket{q_1}$ \\ 
\hline
$\ket{0}$ & $\ket{0}$ & $\ket{0}$ & $\ket{0}$ \\
$\ket{0}$ & $\ket{1}$ & $\ket{0}$ & $\ket{1}$ \\
$\ket{1}$ & $\ket{0}$ & $\ket{1}$ & $\ket{1}$ \\
$\ket{1}$ & $\ket{1}$ & $\ket{1}$ & $\ket{0}$ \\
\hline\hline          
\end{tabular}
\end{center}
\caption{
\label{tab:CNOT} 
Logical table of CNOT gate.}
\end{table}
Rotation and CNOT gates are elementary gates which are composed to {\em quantum circuits} that are required for the input of the classical data into a quantum algorithm as well as for the unitary evolution of the same. The quantum version of the reservoir is composed of exactly these gates. The readout of information is done by a measurement process which causes the collapse of the $n$-qubit quantum state.
\bibliographystyle{unsrt}
\bibliography{references}

\end{document}